\documentclass[11pt]{article}
\usepackage[utf8]{inputenc}
\usepackage{amsthm}
\usepackage{amsmath} 
\usepackage{pgfplots} 
\usepackage{tikz}
\usepackage{sgame} 
\pgfplotsset{compat=1.7}
\usetikzlibrary{intersections, calc}
\usetikzlibrary{patterns, fillbetween}
\usepackage[english]{babel} 
\usepackage{blindtext}
\usepackage{natbib}
\usepackage{amsfonts}
\usepackage{bbm}

\usepackage{comment}
\usepackage[textwidth=30mm]{todonotes}

\newtheorem{theorem}{Theorem}

\newtheorem{corollary}{Corollary}
\newtheorem{definition}{Definition}
\newtheorem{lemma}{Lemma}
\newtheorem{obs}{Observation}
\newtheorem{remark}{Remark}

\def\q1{\tilde{q_1}}
\def\q2{\tilde{q_2}}
\def\e1{\mathbf{e}} 
\def\O{O}

\title{The Indoctrination Game\thanks{For their valuable comments, the authors thank Tomer Blumkin, Ran Eilat,  Ehud Lehrer, Igal Milchtaich, Abraham Neyman, Dov Samet, Aner Sela, and Yevgeny Tsodikovich as well as the participants of the  Tel Aviv University Game Theory and Mathematical Economics Research Seminar, the $12^{\rm th}$ annual conference of the Israeli Chapter of the Game Theory Society, and the BIU Game \& Economic Theory Seminar.}}
\author{Lotem Ikan\thanks{Department of Economics, Ben-Gurion University of the Negev, Israel. E-mail: \textsf{ikanl@post.bgu.ac.il}} \  and David Lagziel\thanks{Department of Economics, Ben-Gurion University of the Negev, Israel.  E-mail: \textsf{Davidlag@bgu.ac.il}}}

\date{\today}

\begin{document}

\maketitle

\thispagestyle{empty}

\begin{abstract}
The indoctrination game is a complete-information contest over public opinion. 
The players exert costly effort to manifest their private opinions in public in order to control the discussion, so that the governing opinion is similar to theirs.
Our analysis provides a theoretical foundation for the silent majority and vocal minority phenomena, i.e., we show that all moderate opinions remain mute in equilibrium while allowing extremists full control of the discussion. 
Moreover, we prove that elevated exposure to others' opinions increases the observed polarization among individuals.
Using these results, we formulate a new social-learning framework, referred to as  \emph{an indoctrination process}.
\end{abstract}

\bigskip

\noindent {\emph{Journal} classification numbers: C72; D62; D72.}

\bigskip


\newpage



\section{Introduction}

The term ``belief" is, to some extent, misused in game theory.
For example, when someone says ``I believe in god", people do not typically assume that this person has a probability distribution (i.e., a belief)  on different states of the world that concern the existence of god, and this distribution is to be updated with newly received information.
There is neither a distribution, nor informative signals in this context.
The same is true for statements such us ``I believe in people's right to X".
These beliefs are a matter of values and preferences, rather than information. 
In fact, you do not need to believe in something that you know, and in game theory, a belief actually reflects \emph{knowledge}.
That is, a ``belief" is the knowledge that certain states may exist, and the knowledge over the probability for each state to be realized. 


This distinction is essential because we have a natural tendency to weave together knowledge with perspectives. 
Consider, for example, the basic structure of what is commonly referred to as a learning model.
We begin with some exogenous uncertainty, a randomly chosen state.
A rational agent then receives new information, typically a signal, and updates his belief accordingly.
This is the fundamental structure of our persuasion models, information-design problems, and  Bayesian-learning processes.
However, in many real-life scenarios, ranging from politics and economics to religion and sports, this fails to accurately reflect the underlying mechanism.
What is the underlying uncertainty in choosing a political side, or in choosing a favorite sports team? 
In fact, two people can mutually agree on all relevant information, and still disagree on the question of who is the greatest football player of all time.
Of course there are some structural uncertainties, but when debating these topics, what we also share are our \emph{opinions}, rather than signals, and opinions work differently.
Thus enters the concept of \emph{indoctrination}.


The indoctrination game is a new type of contest in which players hold fixed private opinions that they discuss with others in what could be described as a public debate.
The players' main goal is to control the discussion, in the sense that the governing opinion is similar to theirs.
More formally, the game comprises a set of individuals whose opinions are distributed on an interval.
These individuals exert costly effort to manifest their opinions in public. 
Their payoffs decrease with the expected distance between their individual opinions and the opinions manifested by others.\footnote{The expected distance it taken in absolute value, so that the weights (i.e., the probabilities) are the players' endogenously generated effort levels.}
That is, opposite opinions do not offset in the players' payoff functions, and they prefer other individuals, whose opinions are far from theirs, to remain silent.
The key ingredient and novelty of this framework is the fact that there is no true state of the world, only different perspectives that collide in equilibrium.


The main goal of this paper is to study the interaction between individuals who hold different opinions, and specifically, the interaction between people who hold moderate opinions and extremists.
This goal is divided into three parts.
First, we study the equilibria of the indoctrination game given that players completely observe the opinions of others.
Next, we analyze a generalized version of the game in which players have limited exposure to others' opinions.
Finally, we use the given results to construct a novel social-learning process in which opinions endogenously evolve in future generations. 
We refer to this dynamic framework as \emph{an indoctrination process}.


To achieve the stated objectives, the paper provides three key results along with several insights.
Our first main result, given in Theorem \ref{Theorem - Linear cost}, establishes a theoretical foundation for the \emph{silent majority} and the \emph{vocal minority} phenomena. 
It shows that moderate opinions remain mute in equilibrium (i.e., the silent majority), while giving extremists the ability to govern the discussion.
Moreover, our analysis indicates that the individuals' inclination to manifest their opinions is inversely related to their level of representation (i.e., the vocal minority).
This negative relation is two-dimensional, depending on the distance between the opinions of the extreme groups, and their sizes.
In other words, once extreme groups reduce in size, or become more extreme, the actions of every remaining individual in these groups intensify on average.
These phenomena were empirically documented by \citet{Mustafaraj2011a} in the context of political discussions on social media.
Their findings were recently supported by a Pew Research Center report, entitled ``National Politics on Twitter: Small Share of U.S.\  Adults Produce Majority of Tweets", which states that $97\%$ of political tweets of U.S.\ adults originate from just $10\%$ of the users, who also hold the least moderate views over the opposing political side.


The intuition behind this crowding-out effect traces back to the augmented relative impact of extreme individuals, one over the other, compared to their impact on moderate players.
Extremists typically try to mitigate the effect of the opposing side, so they naturally exert a higher level of effort on aggregate.
This aggressive behaviour dilutes the impact of moderate players, thus creating a positive feedback loop that intensifies the extremists' behaviour.
The effect eventually stabilizes once all moderate players withdraw from the debate.
This result holds independently of the number of players and opinions.


The second main result relates to the extended model in which players only have partial monitoring over others.
In this set-up, we study how the exposure level of individuals to others' opinions affects the equilibria of the game.
Our analysis shows that an elevated exposure level increases polarization.
To see this, we adapt the seminal polarization metric of \citet{Esteban1994} to our setting, and show that polarization increases in any equilibrium, as a function of the exposure level.
Interestingly, this phenomenon was also empirically documented in a recent field experiment by \citet{Bail2018},  who use politically leaning bots on social media to show that exposure to opposing views increases political polarization.


Focusing on the first two parts of the paper, it is clear that the act of indoctrination, within the given framework, is rather futile. 
The players' main objective, as indicated by their payoff functions, is to influence others by controlling the discourse. 
However, in this one-stage setting, players do not alter their opinions. 
This issue is addressed in the third part of this paper, which delves into a new evolutionary, adaptive-learning framework.


In the third part of this paper, we use the equilibrium result of the limited-exposure model to \emph{endogenously} generate a transition matrix between opinions that yields an inter-generational adaptive-learning process. 
Specifically, in every stage, individuals act according to some equilibrium profile, and the distribution of opinions of the subsequent generation is determined in proportion to the observed opinions given that profile. 
This generates a non-Bayesian evolutionary process where the transition matrix is repeatedly derived according to the newly realized equilibrium. 
Our analysis focuses on the stationary distribution of the learning process as a function of the exposure level, and shows that a higher exposure level leads to a more polarized society. 
In other words, we demonstrate that the distribution of opinions spreads further apart as the exposure level increases.\footnote{Note that in this context, \emph{"social learning"} refers to the cognitive and evolutionary process of observing and absorbing the subjective opinions of others and should not be confused with either Bayesian or other forms of rational learning.}





\subsection{Relation to literature}

Our basic framework lays in the vast literature of contests which goes back to the seminal study of \citet{Tullock1980}, and later followed by \citet{Skaperdas1996}  and \citet{Baye2003}, among many others.
Within this set of games, there exists a specific class of \emph{contests with externalities},\footnote{This feature ranges also to auction theory,  which accommodates a vast literature on identity-dependent externalities; see, for example, \citet{Funk1996}, \citet{Jehiel1996a,Jehiel1999}, \citet{Varma2002}, \citet{Aseff2008},  and \citet{Brocas2013}. } motivated by the early work of \citet{Buchanan1980}, and more substantially by the later work of \citet{Congleton1989} which studies status-seeking contests with externalities that affect outside (non-strategic) individuals.
In recent years, this research area expended in various directions,\footnote{See, e.g.,  \citet{Chung1996}, \citet{Lee1998}, \citet{Eggert2006}, \citet{Shaffer2006},   \citet{Konrad2006}, \citet{Lee2007},  \citet{Cohen2008},  \citet{Chowdhury2011}, \citet{Ahn2011}, \citet{Klose2015}, and \citet{Park2019}, among many others.} thus we shall focus on studies that are closest to the current research agenda.

The early studies of Tullock contests were generalized by \citet{Linster1993}, that derives an equilibrium in pure strategies in a setting where losing players  are not indifferent to the identity of the winner.
Although our framework focuses on different payoff functions,  the linear cost function allows us to use similar mathematical methods as the ones used by \citet{Linster1993}. 
Another key feature of our setup goes back to the work of \citet{Nitzan1991}, which studies Tullock contests where players are partitioned into groups who compete together. 
Once a group wins the prize,  they apply various sharing rules to divide the prize among its members.
The concept of partitioning players into competing groups is quite natural in the context of public debates,  and would indeed prove important in our setup, as well.

Similarly to \citet{Moldovanu2012} and \citet{Sela2020},  the indoctrination game also encompasses negative externalities.
In our framework however all payoffs are negative,  rather than a combination of prizes and penalties, carrots and sticks.
In general,  contests with externalities could also be classified according to the type of externalities and the individuals that are affected by them.
The indoctrination game falls within the  set of contests with negative, identity-dependent externalities that affect all players, independently of their winning status.

Overall, the study that is closest to the first two parts of this paper is the seminal work of \citet{Esteban1999}, and specifically Section $5$ therein.
Our basic model extends \citet{Esteban1999}, by generalizing the payoffs and groups of players (using the ``linear alienation'' given in \citealp{Esteban1994}) and by focusing on different cost functions (similarly to \citealp{Linster1993}).
Evidently our results give rise to additional conclusions, the obvious one being that the silent-majority and vocal minority phenomena,  and the emergence of the stated crowding-out effect in equilibrium.


The third part of our study, which builds on the first two parts, lays in the intersection of adaptive learning and evolutionary processes. 
Our analysis, however, is closer to the former rather than the latter.
The concept of adaptive learning can be traced back to the work of \citet{DeGroot1974}, which investigates the stochastic process of reaching a consensus by adapting observed opinions. 
Our motivation and general objective are quite similar to this line of research. 
In many of these studies, players follow certain heuristics, such as Na\"ive learning as in \citet{Golub2010} and \citet{Amir2021a}, or the  majority dynamics as in \citet{Galam2002} and \citet{Arieli2023},  that do not necessarily establish an equilibrium in the relevant framework.
Our analysis offers a different perspective in two fundamental issues: first, we base the learning process on the equilibria of the limited-exposure indoctrination game; and second, there is no true state of the world in our setting, only opinions.
Notably, this allows us to link contest theory with social learning while providing a micro-founded framework for non-Bayesian adaptive processes.



\section{The game} \label{Section - The game}

The indoctrination game is a complete-information, single-stage contest in which players hold fixed individual opinions that they manifest in public.
To do so, the players exert costly effort and are being rewarded according to the distance between the aggregate distribution of publicly observed opinions and their private ones.
In equilibrium, players balance their individual cost of effort with the need to shift the public opinion towards their own.


Formally, fix $k\geq 2$ distinct values $\O_1 < \O_2 <  \cdots < \O_k$ in $\mathbb{R}$, that represent $k$ different opinions.
We shall refer to $\O_1$ and $\O_k$ as the \emph{extreme opinions}, and to all others as \emph{moderate} ones.\footnote{To facilitate the exposition, we sometimes relate to players with extreme/moderate opinions as extreme/moderate players,  respectively.}
Let $N=\{1,2,\dots,n\}$ be the set of players, and for every $i=1,\dots,k$, let $N_i$ denote the non-empty set of players with a private opinion $\O_i$, such that $n_i= |N_i| \geq 1$ and $n =\sum_i n_i$. 
We refer to the players in $N_i$ as the $\O_i$-players.


The action set of every player is $\mathbb{R}_+$.
An action $e_j \geq 0$ is the effort that player $j \in N_i$ exerts to publicly manifest his opinion $\O_i$. 
Given a non-zero action profile $\e1 =(e_{1},\dots,e_{n}) \in 
 \mathbb{R}_+^n$, consider the random variable $X_{\e1}$ distributed according to
$$
\Pr(X_{\e1}=\O_i) = \frac{\sum_{j \in N_i} e_j }{\sum_{j=1}^n e_j} = \frac{E_i }{\sum_{j=1}^k E_j},
$$
where $E_i = \sum_{j\in N_i} e_j$ is the sum of efforts of all $\O_i$-players.
Intuitively, $P_{X_{\e1}}(\cdot)$ is the distribution of publicly observed opinions,  weighted according to the players' effort levels.
If, for example, all $\O_i$-players exert relatively high effort levels (on aggregate and compared to all other players combined), then their opinion would dominate the debate and $X_{\e1}$ would be distributed accordingly.



The expected payoff of player $j\in N_i$, given a non-zero effort profile $\e1 \in \mathbb{R}^n_+$, is 
\begin{equation*}
    U_j(\e1|\O_i) = -e_j - \mathbb{E}[|\O_i - X_{\e1}|].
\end{equation*} 
The payoff function presents the classic tension in contest theory between the private cost of effort $e_j$ and the need to govern the debate.
The term $\mathbb{E}[|\O_i - X_{\e1}|]$ is the expected distance between opinion $\O_i$ and publicly observed opinions, given the players' effort levels $\e1$. 
Thus, in case the distribution of  publicly observed opinions $X_{\e1}$ shifts towards $\O_i$, then all $\O_i$-players benefit from the reduced expected distance $\mathbb{E}[|\O_i - X_{\e1}|]$.
Note that the expected distance is taken in absolute value, so opposing opinions (relative to $\O_i$) do not offset.
To eliminate trivial results of a null debate in which no player exerts positive effort (i.e., to exclude $e_0 =(0,0,\dots,0)$ as an equilibrium), fix $U_j(e_0|\O_i)=\inf_{{\e1}\in \mathbb{R}_+^n \setminus\{e_0\}} U_j({\e1}|\O_i)$ for every opinion $\O_i$ and for every player $j$.\footnote{Nash equilibria are robust to affine payoff transformations, so if needed, one can adjust the payoff functions to get strictly positive payoffs under undominated strategies.}


\section{The silent majority and the vocal minority} \label{Section - The silent majority and the vocal minority}

Our analysis begins with equilibria characterization.
Theorem \ref{Theorem - Linear cost} describes the equilibria of the indoctrination game, and doing so, presents two intriguing phenomena.
The first, referred to as the \emph{silent majority}, shows that all moderate players, i.e., players who do not posses extreme opinions, remain silent in every equilibrium. 
The theorem formally states that, in every equilibrium, the effort level of every moderate individual is zero.
In other words, the only players who extract positive effort levels in equilibrium are the ones who hold the extreme opinions $\O_1$ and $\O_k$.\footnote{We acknowledge that the majority could be based in the extremes. The terminology relates to the typical case in which the extremists are relatively small groups.}

The second phenomenon, which complements the first, is referred to as the \emph{vocal minority}.
Not only that the extreme opinions completely govern the public debate, the average expected effort of every individual in these groups is inversely related to theirs sizes.
In other words,  individuals from smaller extreme groups tend to be louder on average.
This follows from the fact that the aggregate effort of each of these groups in equilibrium depends solely on the distance $|\O_1 - \O_k|$. 
So if one group is smaller than the other, the average ``vocality" (i.e., effort level) of every individual in the smaller group is higher.
Before presenting Theorem \ref{Theorem - Linear cost}, we emphasize that the results are independent of the relative position of opinions and the number of moderate players.
This underscores the robust nature of the two aforementioned phenomena.

\begin{theorem} \label{Theorem - Linear cost}
In every equilibrium,  the effort level of every moderate player is zero,  whereas the aggregate effort levels of all extreme players are $E_1 = E_k = \tfrac{|\O_1-\O_k|}{4}$.
\end{theorem}

An immediate corollary, following Theorem \ref{Theorem - Linear cost}, relates to the unique symmetric equilibrium in which every extreme player exerts the same level of effort as all other players sharing the same opinion. 
(The proof is follows immediately from Theorem \ref{Theorem - Linear cost}, thus omitted.)

\begin{corollary} \label{Corollary - Uniqe sym. eq.}
There exists a unique symmetric equilibrium $ {\e1}^{\rm sym}$ such that 
\begin{equation*}
    {\e1}^{\rm sym}_j = \begin{cases}
        0, & \text{ }  \forall j\in N_i, i\neq 1,k, \\
        \tfrac{|\O_1-\O_k|}{4n_i}, & \text{ } \forall j\in N_i, i=1,k,
        \end{cases}
\end{equation*}
and the expected payoff of every player $j$, given $ {\e1}^{\rm sym}$, is
\begin{equation*}
U_j({\e1}^{\rm sym}|\O_i) = -\frac{|\O_1-\O_k|}{2} \cdot \left[ 1 + \frac{1}{2n_i}\mathbbm{1}_{\{i=1,k\}}\right].
\end{equation*}
\end{corollary}

The driving force and intuition behind this result is the \emph{crowding-out} effect of extreme players over moderate ones in equilibrium.
The impact of extreme players from both sides, one over the other, is significantly higher than their impact on moderate players (in proportion to the distance between the different opinions). 
So extreme individuals naturally aim to mitigate the effect of other extreme players by increasing their effort levels.
This joint ``aggressive" behaviour dilutes all other opinions (note that the denominator in $P_{X_{\e1}}(\cdot)$ becomes larger), so individuals with moderate opinions are less inclined to extract effort, thus producing a positive feedback loop which results in the stated equilibrium.
This is a somewhat extensive, yet natural, \emph{crowding-out} effect in equilibrium.
The effect stabilizes once all moderate opinions withdraw from the public debate, whereas the aggregate effort levels of the extreme individuals adjust to $\tfrac{1}{4}|\O_i-\O_k|$.

There are several additional conclusions that one can derive from Theorem \ref{Theorem - Linear cost}: (i) The crowding-out effect is beneficial for moderate players who retain a strictly higher expected payoff, compared to extremists. Moderate individuals actually increase their payoff by not participating in the public debate, whereas extreme players are bound to invest heavily in this contest; (ii) Everyone lose from polarization. The expected payoffs of all players are proportional to $|\O_1-\O_k|$, so additional separation between extreme opinions is  detriment. Moreover, extreme players lose the most from polarization; (iii) Free-riding may originate in equilibrium within each group of extreme players. The aggregate effort levels of extreme individuals are independent of the groups' sizes, so extreme players benefit from the participation of others extremists within the same group. This is supported by Corollary \ref{Corollary - Uniqe sym. eq.} which shows that, under the unique symmetric equilibrium,  the expected payoff of extreme players increases with their groups' sizes; and (iv) The equilibria of the game are independent of the relative position and of the number of moderate players.  In other words, the relative position of the polarized groups is the key factor that ``sets the tone" in the debate. Yet, we stress that this result may change if we divert from a linear cost function, specifically to either convex, or concave cost functions.

\section{Limited exposure in public debates} \label{Section - Limited exposure in public debates}

The basic indoctrination game builds on the premise of full monitoring, i.e., that individuals fully observe the opinions of all others.
In practice, however, the exposure and attention of players vary, so one should also consider the possibility of a partial-monitoring setting in which players have limited exposure to others' opinions.
These limitations could arise from external reasons such as network effects, as well as internal ones, e.g., to preempt cognitive inconsistencies.
Namely, when people only partially agree with some ideas, they may refrain from spreading them, thus affecting the ability of others to observe these ideas.
Moreover, even if some opinions eventually do become public, people may feel an internal urge to partially ignore them, specifically because they do not match their private ones.

In this section we study how limited exposure/attention to opposing views impacts visible polarization in the debate.\footnote{To simplify the exposition, we follow the limited-exposure terminology in this section, but one could similarly interpret all results to limited attention.}
Our results show that, at least in the short term (i.e., as long as opinions do not shift), an elevated exposure to opposing opinions has an adverse effect on polarization, making the debate more intense.
To gain some preliminary intuition for this statement, consider splitting the basic indoctrination game (given in Section \ref{Section - The game}) into two separate games, each with at least two opinions, so that the first contains all players with opinions $\O_1$ through $\O_{\lfloor k/2 \rfloor}$, and the second contains all players with opinions $\O_{\lfloor k/2\rfloor +1}$ through $\O_{k}$. 
Theorem \ref{Theorem - Linear cost} predicts that  the extreme individuals within each of these sub-games would control the discussion in proportion to $|\O_{1} - \O_{{\lfloor k/2 \rfloor}}|$ and $|\O_{{\lfloor k/2 \rfloor}+1} - \O_{k}|$, respectively. 
In other words, the fragmentation into two separate sub-games reduces the (internal) intensity within each debate.
Thus, the reverse procedure through which distinct sub-groups better observe each other,  evidently generates a high-intensity debate in equilibrium. 
This provides some intuition for the conclusion that the debate intensifies the more people are exposed to others' opinions, and it also provides a conceptual framework for the recent empirical evidence provided by \citet{Bail2018} who show how exposing people to opposing views in social media increases political polarization.
To formally discuss and prove these statements, we first define a \emph{limited-exposure} indoctrination game, and then adjust the polarization metric of \citet{Esteban1994} to our context.

To capture the notion of partial monitoring, we introduce an \emph{exposure level} $\delta \in (0,1]$ which limits the ability of players to observe distant opinions.
More formally, consider the previously defined indoctrination game, but assume that a fraction of the information that a $\O_l$-player generates is discarded, by a factor of $\delta^{|i-l|}$, until it reaches a $\O_i$-player. 
In such a case, the payoff function of every player $j\in N_i$ takes the following form 
\begin{eqnarray*}
    U_j({\e1}|\O_i) 
    & = & -e_j - \frac{\sum_{l=1}^k E_l \delta^{|\O_i-\O_l|}|\O_l - \O_i| }{\sum_{l=1}^k \delta^{|\O_i-\O_l|} E_l}.
\end{eqnarray*}
In simple terms, the players' exposure to each other decreases as a function of the distance between their individual opinions.

\begin{remark}
Before we elaborate on the polarization metric, let us clarify that the analysis
throughout this section is confined to a symmetric set-up with three opinions, i.e., $k=3$ and $|\O_1-\O_2|=|\O_2-\O_3|=1$.
This assumption is imposed for tractability, and the analysis of the general case, with any number of opinions and valuations, is left for future research.
We refer to this limited framework as the \emph{limited-exposure} indoctrination game.
\end{remark}

To measure polarization in public debates, we follow the seminal work of \cite{Esteban1994} that axiomatically construct the following polarization metric for populations with various characteristics (see Theorem $1$ and $2$, as well as Section $5.1$, therein). 
We adopt their metric by taking $E_i$ to be the observed volume of opinion $i$, so that the effort profile ${\e1}\in  \mathbb{R}_+^n$ translates to a polarization value of
\begin{equation}
    P({\e1}) = \frac{\sum_{i,j} E_i^2E_j |\O_i - \O_j|}{\left[\sum_i E_i\right]^3}.
\end{equation}
This polarization metric is invariant to the aggregate volume of opinions, and typically increases once masses are shifted towards the extremes (see Axioms $1-3$ and Condition H in \citealp{Esteban1994}).
Notably, the result given in Theorem \ref{Theorem - Linear cost} above supports the highest possible level of polarization in the general case (of $k$ opinions).\footnote{See Theorem $2$ in \cite{Esteban1994} concerning the bimodal distribution.}

The polarization level $P({\e1})$ clearly depends on the induced profile ${\e1}\in \mathbb{R}^n_+$ in equilibrium, which in turn depends on the exposure level $\delta$.
So, any discussion about polarization must first specify the equilibrium profile ${\e1}$.
For this purpose, we take the broad objective of considering the impact of the exposure level on \emph{all} possible equilibria. 
Formally, 
\begin{definition}
let $\Lambda(\delta)$ be the set of all equilibria in the limited-exposure indoctrination game with exposure level $\delta$.
We say that the polarization in the limited-exposure indoctrination game \emph{increases} in its exposure level if $P({\e1}_1) > P({\e1}_2)$, for every $\e1_1 \in \Lambda(\delta_1)$, every $\e1_2 \in \Lambda(\delta_2)$, and every $\delta_1 > \delta_2$.
\end{definition} 
\noindent In other words, we do not restrict our analysis through some equilibrium selection, but consider all possible equilibria of the limited-exposure game.

Our main result in this section, given in Theorem \ref{Theorem - Polarization increase} below, indeed shows that the  polarization in a given game increases in its exposure level.
The intuition behind this result is the augmented relative impact of extreme players on each other, relative to their impact on moderate players.
Once the exposure increases, the relative impression of extreme players on each other becomes significant, so that they manifest their opinions more strongly, thus diluting the impression of all moderate players and making the polarization evident.

\begin{theorem} \label{Theorem - Polarization increase}
    The polarization level of the limited-exposure game strictly increases in its exposure level.
\end{theorem}

To prove Theorem \ref{Theorem - Polarization increase} we require the following supporting lemma which states that, in any equilibrium, moderate players become relatively less vocal once the exposure increases.

\begin{lemma} \label{Lemma - Ratio of moderate to extreme}
    For any given exposure level, the ratio between the aggregate effort level of moderate players and that of extreme players, in every equilibrium, is unique and strictly decreases in $\delta$.
\end{lemma}

\begin{figure}[ht] 
\centering
\begin{tikzpicture}

\begin{axis}[
axis x line=middle, 
axis y line=middle,
xlabel={$\delta$}, 
ylabel={$\frac{E_2}{E_1+E_3}$}, 
xmin=-0.1, 
xmax=1.1, 
ymin=-0.05, 
ymax=0.72, 
xtick={0,0.2,0.4,0.6,0.8,1},
ytick={0,0.3,0.6},
domain=0.02:1, 
samples=100, 
]
\addplot[color=blue,mark=none,thick]{1/3*(x^2-3*x)+ 1/6 * (8*x^6 - 36*x^5 + 27*x^4 + 36*x^3 - 54*x^2+3*sqrt(3)*sqrt(-8*x^9 + 27*x^8 + 24*x^7 - 108*x^6 - 24*x^5 + 162*x^4 + 8*x^3 - 108*x^2 + 27) + 27)^(1/3)-(-16*x^4 + 48*x^3 - 48*x) / (24*(8*x^6 - 36*x^5 + 27*x^4 + 36*x^3 - 54*x^2+ 3*sqrt(3)*sqrt(-8*x^9 + 27*x^8 + 24*x^7 - 108*x^6 - 24*x^5 + 162*x^4 + 8*x^3 - 108*x^2 + 27) + 27)^(1/3))}; 
\end{axis}
\end{tikzpicture}\caption{\footnotesize The ratio between the aggregate effort levels of moderate players to  that of extreme players, in equilibrium, as a function of the exposure level. Though the equilibrium is not unique, the relation between $\tfrac{E_2}{E_1+E_3}$ and $\delta$ does hold in every equilibria of the limited-exposure $3$-player indoctrination game.} \label{Figure - Ratio Agg. effort levels and exposure.}
\end{figure}
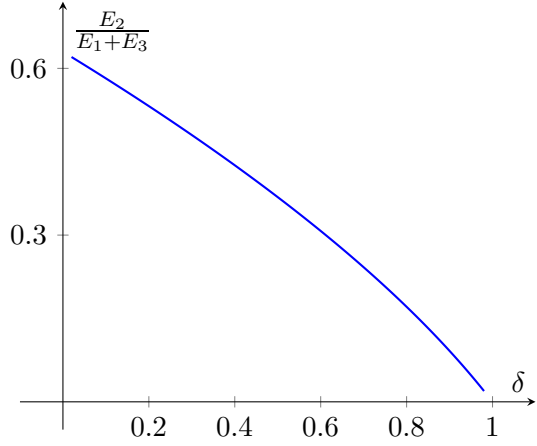

Figure \ref{Figure - Ratio Agg. effort levels and exposure.} depicts the functional relation, described in Lemma \ref{Lemma - Ratio of moderate to extreme},  between $\tfrac{E_2}{E_1+E_3}$ and the exposure level $\delta$ in any equilibrium of the limited-exposure game.
The relation is implicitly given by the following equation 
$$
4\left(\delta + \tfrac{E_2}{E_1+E_3}\right)^3 = \left[ 1+\delta^2 + 2\delta \tfrac{E_2}{E_1+E_3} \right]^2,
$$ 
as derived in the proof of Lemma \ref{Lemma - Ratio of moderate to extreme}.
In case $\delta$ tends to $1$, one can see that we converge to the baseline model studied in the previous section, so that the ratio $\frac{E_2}{E_1+E_3}$ tends to zero in equilibrium.

\section{Dynamic opinions: an indoctrination process} \label{Section - Dynamic opinions}

The limited-exposure indoctrination game allows us to discuss, at least in general terms, the possibility of dynamic opinions.
Consider, for example, the basic majority-rule (reaction-diffusion) model as in \citet{Galam2002}, in which people are repeatedly and randomly matched into subgroups, so that in every stage, each individual adapts the opinion of the majority within his group. 
To some extent, this is a reduced-form non-strategic model of indoctrination, in which people simply conform to the opinions of others.

To extend this model to our strategic setting, we propose an updated framework called the \emph{indoctrination process}, which involves two adjustments.
First, instead of assuming a fixed set of individuals, we consider an inter-generational process where players are replaced in every stage.
Second, instead of using the majority rule, we determine the opinions of the newly formed players in each stage based on the distribution of opinions and equilibrium profile from the previous stage.

More formally, for every $\delta \in (0,1]$ and for every stage $t\geq 0$, denote by $\e1^t$ an equilibrium profile of the limited-exposure game (assuming that all opinions are represented), and consider the $3\times 3$ transition matrix $Q^t$ with entries $Q^t_{i,j}= \Pr(X_{\e1^{
t}} = \O_j|O_i)$.
Explicitly, 
$$
Q^t =
\begin{bmatrix}
\frac{E_1}{E_1+\delta E_2 +\delta^2 E_3} & \frac{\delta E_2}{E_1+\delta E_2 +\delta^2 E_3} & \frac{\delta^2 E_3}{E_1+\delta E_2 +\delta^2 E_3} \\
\frac{\delta E_1}{\delta E_1+ E_2 +\delta E_3} & \frac{E_2}{\delta E_1+E_2 +\delta E_3} & \frac{\delta E_3}{\delta E_1+ E_2 +\delta E_3} \\
\frac{\delta^2 E_1}{\delta^2 E_1 +\delta E_2 + E_3} & \frac{\delta E_2} 
{\delta^2 E_1 +\delta E_2 + E_3} & \frac{E_3}{\delta^2 E_1 +\delta E_2 + E_3} \\
\end{bmatrix}.
$$
We use this matrix structure to define the following dynamic process.
In stage $t=0$, the players' opinions are fixed according to some initial distribution $\pi_0$ with full support. These players act according to an equilibrium profile $\e1^0$.
In stage $t=1$, a new generation is formed, and their opinions are distributed according to $\pi_1=\pi_0 Q^{0}$, where $Q^0$ is the previously defined transition matrix associated with $\e1^0$.
In simple terms, the generation in stage $t=1$ observes the public opinion generated by the previous generation, which depends both on the equilibrium profile $\e1^0$ and on the initial distribution  $\pi_0$.
Subsequently, in each stage $t\geq 1$, the newly formed generation adapts the opinion distribution $\pi_t$ according to the following equation: $\pi_t=\pi_{t-1}Q^{t-1}$, where $Q^{t-1}$ is the transition matrix associated with the equilibrium $\e1^{t-1}$. 
This process continues indefinitely.\footnote{If $\pi_t$ contains irrational values, it will not be feasible to implement it with a finite set of players. In such cases, one can use a sufficiently close approximation of $\pi_t$, which would also yield sufficiently close results. The notion of $M$-absorbing sets, as discussed in \cite{Lehrer2021}, is helpful in this regard.}

The indoctrination process works such that a newly formed generation observes the opinions of the previous one in equilibrium, while taking into account the different perspectives of each subgroup. 
This process builds on an inherent biased, as the previous distribution of opinions can significantly influence the subsequent one through the observed opinions. 
For instance, if the newly formed generation belongs to a population that is heavily skewed in favor of a particular opinion, say $O_1$, then their opinions would be significantly influenced by the viewpoints of $O_1$-players in equilibrium.
Now, we can use the generic equilibrium profile given in the proof of Lemma \ref{Lemma - Ratio of moderate to extreme} to explicitly present the transition matrix in every stage $t$.

\begin{obs}
The transition matrix in every stage $t$ and in every equilibrium $\e1^t$ \emph{(}as given in the proof of \emph{Lemma \ref{Lemma - Ratio of moderate to extreme}}\emph{)} is
$$
Q^t =
\begin{bmatrix}
\frac{1}{1+\delta r^* +\delta^2 } & \frac{\delta r^*}{1+\delta r^* +\delta^2 } & \frac{\delta^2}{1+\delta r^* +\delta^2 } \\
\frac{\delta }{2\delta +r^*} & \frac{r^*}{2\delta +r^*} & \frac{\delta }{2\delta +r^*} \\
\frac{\delta^2 }{1+\delta r^* +\delta^2} & \frac{\delta r^*}{1+\delta r^* +\delta^2} & \frac{1}{1+\delta r^* +\delta^2}
\\
\end{bmatrix},
$$
where $r^* = \frac{E_2}{E_1}$.
\end{obs}
Note that this is a right centrosymmetric transition matrix, i.e., it is symmetric with respect to its center $Q^t_{2,2}$ and every row sums to one.
Moreover, as long as all opinions are represented, the ratio $r^* = \frac{E_2}{E_1}$ is independent of the number of players holding each opinion.
So,  for every $\delta \in (0,1)$, this irreducible and aperiodic transition matrix holds in every stage $t$ and in every equilibrium $\e1^t$.
Thus, the convergence towards its unique, stationary, probability eigenvector $\pi$ is guaranteed independently of the initial distribution of opinions.
Specifically, its stationary distribution is 
$$
\pi = \left( \frac{\sqrt{\delta +\tfrac{1}{2}r^*}}{2\sqrt{\delta +\tfrac{1}{2}r^*}+r^*}, \frac{r^*}{2\sqrt{\delta +\tfrac{1}{2}r^*}+r^*} ,\frac{\sqrt{\delta +\tfrac{1}{2}r^*}}{2\sqrt{\delta +\tfrac{1}{2}r^*}+r^*}\right).
$$

Lemma \ref{Lemma - Ratio of moderate to extreme} states that $r^*$ is a decreasing function of $\delta$, so one can easily prove that $\pi_2$ is decreasing in $\delta$ as well, thus establishing that the population becomes more polarized as $\delta$ increases.
\begin{lemma} \label{Lemma - polarized population}
    The proportion $\pi_2$ of moderate players  decreases in $\delta \in (0,1]$.
\end{lemma}

Besides monotonicity, we can use the functional relation between $\delta$ and $r^*$, given after Lemma \ref{Lemma - Ratio of moderate to extreme}, to compute the stationary distribution in case $\delta$ tends to either $0$ or $1$.
Specifically, in case $\delta$ tends to $0$, the stationary distribution converges to $\pi=\left( \tfrac{1}{2+2^{1/3}},\tfrac{2^{1/3}}{2+2^{1/3}} ,\tfrac{1}{2+2^{1/3}}\right) \approx  (0.307,0.386,0.307)$.
On the other hand, in case $\delta$ tends to $1$, we know that $r^*$ converges to $0$, and so we get $\pi=(0.5,0,0.5)$ in case of full exposure. 
In other words, if there are no limitations and everyone can observe all opinions, the entire population reaches the most extreme state of polarization.

\section{In conclusion}

The indoctrination game offers a valuable perspective on social debates, which goes beyond the formal results presented in this paper.
It presents an alternative framework to the standard Bayesian inference and rational-learning models, allowing for players to indoctrinate each other. 
This shift in perspective prompts a reevaluation of the assumption that there is always an objective, unknown state of the world that individuals seek to discover. 
Instead, it recognizes the possibility that people may hold differing opinions based on their subjective life experiences. 
The game provides a theoretical foundation for empirically documented phenomena such as the silent majority and vocal minority, as well as the impact of exposure to opposing opinions on polarization within a population. 
However, the game should not be regarded as a restrictive approach to social debates, but rather as an alternative framework that allows for a more nuanced understanding of how people form and revise their beliefs in social settings.

\newpage

\bibliographystyle{aer}
\bibliography{./references}

\newpage

\appendix

\section{Proof of Theorem \ref{Theorem - Linear cost}}

\begin{proof}
The zero vector is clearly not an equilibrium, so fix a non-zero profile ${\e1} \in \mathbb{R}^n_+$, and consider the payoff function of player $j\in N_i$, 
\begin{eqnarray*}
    U_j(e_j,e_{-j}|\O_i) 
    & = & -e_j - \frac{\sum_{l=1}^k\sum_{r \in N_l} e_r|\O_l - \O_i| }{\sum_{r=1}^n e_r} \\
    & = & -e_j - \frac{\sum_{l=1}^k E_l |\O_l - \O_i| }{\sum_{l=1}^k E_l}.
\end{eqnarray*}
The function $U_j(\cdot,e_{-j}|\O_i)$ is differentiable and concave in $e_j$, so the maximum is reached either at the boundary $e_j=0$ (effort levels are unbounded from above), or when the following FOC is satisfied:
\begin{equation*} 
    \frac{\partial U_j(e_j,e_{-j}|\O_i)}{\partial e_j} = \sum_{l=1}^kE_l|\O_l - \O_i| - \left[ \sum_{l=1}^k E_l \right]^2=0, \ \ \forall j=1,\dots,n.
\end{equation*}
Denote $d_{l,i} = |\O_l - \O_{i}|$, and note that
$$
d_{l,i} - d_{l,i+1} = |\O_l - \O_{i}|-|\O_{l} - \O_{i+1}|=
\begin{cases}
    - d_{i,i+1}, & \ \ \forall l\leq i, \\
    d_{i,i+1}, & \ \ \forall l> i.
\end{cases}
$$
For every $i=1,\dots,k-1$, compute the difference 
\begin{eqnarray}
    \frac{\partial U_j({\e1}|\O_i)}{\partial e_j} - \frac{\partial U_{j'}(e|\O_{i+1})}{\partial e_{j'}} 
    & = & \sum_{l=1}^kE_l d_{l,i} - \sum_{l=1}^kE_l d_{l,i+1} \nonumber \\
    & = & -\sum_{l \leq i}E_ld_{i,i+1} + \sum_{l > i}E_ld_{i,i+1} =0. \label{Equation - FOC}
\end{eqnarray}
Divide every such Equation \ref{Equation - FOC} (for opinion $\O_i$) by $d_{i,i+1}\neq 0$ to get
$$
H_i := -\sum_{l \leq i}E_l+ \sum_{l \geq  i+1}E_l =0.
$$
Subtract $H_{i-1} - H_{i}$ to get $2 E_i= 0$ for every $i=2,\dots,k-1$.
Since effort levels are non-negative, we deduce that, in equilibrium, the first-order conditions are satisfied at the boundary $e_j=0$, for every moderate player $j$. Thus, we are left with the following FOCs for the  extreme opinions
\begin{eqnarray*}
        E_i |\O_1-\O_k| - \left[ E_1 + E_k \right]^2=0, \ {\rm where \ } i=1,k.
\end{eqnarray*}
Solving for $E_1$ and $E_k$, we get a unique solution (other than the zero-effort profile) of $E_1=E_k = \frac{|\O_1-\O_k|}{4}$, as needed.
\hfill
\end{proof}

\section{Proof of Theorem \ref{Theorem - Polarization increase}}

\begin{proof}
Consider an equilibrium profile ${\e1} \in \mathbb{R}^n_+$. 
It follows from the proof of Lemma \ref{Lemma - Ratio of moderate to extreme} that $E_1=E_3$, so the polarization level translates to
\begin{eqnarray*}
    P({\e1}) 
    & = & \frac{2E_1E_2^2 +  2E_1^2E_2 +4E_1^3 }{\left[2E_1 + E_2\right]^3} = \frac{  W^2 + \tfrac{1}{2}W + \tfrac{1}{2}}{\left[1+W\right]^3} = \frac{1}{1+W} - \frac{1}{2(1+W)^2}-\frac{W}{(1+W)^3},
\end{eqnarray*}
where $W=\frac{E_2}{2E_1}$.
According to Lemma \ref{Lemma - Ratio of moderate to extreme}, $W$ is strictly decreasing in $\delta$, so it is left to prove that $P({\e1})$ is strictly decreasing w.r.t.\ $W\geq 0$.
Evidently,
$$
\frac{d P}{d W} = -\frac{1}{(1+W)^2} + \frac{3W}{(1+W)^4} = \frac{-W^2+W-1}{(1+W)^4} <0,
$$
for every $W\geq 0$, as needed.
\hfill \end{proof}

\section{Proof of Lemma \ref{Lemma - Ratio of moderate to extreme}}

\begin{proof}
Fix $\delta \in (0,1)$ and consider the FOCs of every player $j \in N_i$ given a non-zero profile $e$, 
$$
\sum_{l=1}^3 E_l \delta^{|i-l|}|\O_l - \O_i| = \left[\sum_{l=1}^3 \delta^{|i-l|} E_l\right]^2,
$$
where $E_l=\sum_{r\in N_l}e_r$ for $1\leq l \leq 3$.
Stated explicitly for every opinion, we get
\begin{eqnarray*}
       {\rm for} \ j \in N_1 & : & \ E_2 \delta + 2 E_3 \delta^2 = \left[ E_1 + E_2 \delta + E_3 \delta^2  \right]^2, \\
       {\rm for} \ j \in N_2 & : & \ E_1 \delta +E_3 \delta = \left[ E_1 \delta + E_2 + E_3 \delta  \right]^2, \\
       {\rm for} \ j \in N_3 & : & \ 2 E_1 \delta^2 + E_2 \delta = \left[ E_1 \delta^2 + E_2 \delta + E_3  \right]^2.
\end{eqnarray*}
Define $X=E_1 + E_2 \delta + E_3 \delta^2$, $Y=E_1 \delta + E_2 + E_3 \delta$, and $Z=E_1 \delta^2 + E_2 \delta + E_3$.
So, 
\begin{eqnarray*}
    X - E_1 + \delta^2 E_3 = X^2, \\
    Y - E_2 = Y^2, \\
    Z - E_3 + \delta^2 E_1  = Z^2.
\end{eqnarray*}
and 
\begin{eqnarray*}
    X-\delta Y = E_1(1-\delta^2) \ \ \Rightarrow \ \ E_1 = \frac{X-\delta Y}{1-\delta^2}, \\
    Z-\delta Y= E_3(1-\delta^2) \ \ \Rightarrow \ \ E_3 = \frac{Z-\delta Y}{1-\delta^2}.
\end{eqnarray*}
Plug this in the previous equations to get
\begin{eqnarray*}
    X^2 = X - \frac{X-\delta Y }{1-\delta^2}  + \delta^2 \frac{Z-\delta Y}{1-\delta^2} =  X + \frac{\delta^2 Z -  X}{1-\delta^2}  + \delta Y  \ \ \Rightarrow \ \  X^2 (1-\delta^2) = (Z-X)\delta^2  + \delta(1-\delta^2) Y, \\
    Z^2 = Z - \frac{Z-\delta Y}{1-\delta^2} + \delta^2 \frac{X-\delta Y}{1-\delta^2} = Z + \frac{\delta^2 X -Z}{ 1-\delta^2} +\delta Y  \ \ \Rightarrow \ \   Z^2(1-\delta^2) = (X-Z)\delta^2+ \delta(1-\delta^2) Y.
\end{eqnarray*}
Subtracting both equations yields $ (X^2-Z^2)(1-\delta^2) +2(X-Z)\delta ^2=0$.
Hence, we conclude that $X=Z$ is the unique solution and $E_1=E_3$.

So, the FOCs revert to 
\begin{eqnarray*}
       2 \delta^2 + 2\delta W =E_1 \left[ 1+\delta^2 + 2\delta W  \right]^2, \\
       2 \delta = E_1 \left[ 2\delta+  2 W   \right]^2,
\end{eqnarray*}
where $W=E_2/(2E_1)$.
Divide the first equation by the second to get
$$
\delta + W = \left[ \frac{1+\delta^2 + 2\delta W}{ 2(\delta + W)} \right]^2 \ \ \ \Leftrightarrow \ \ \ 4(\delta + W)^3 = \left[ 1+\delta^2 + 2\delta W \right]^2.
$$
Define the function $Q(W,\delta) = 4(\delta + W)^3 -\left[ 1+\delta^2 + 2\delta W \right]^2$ 
and note that $Q(0,\delta)<0$ and $Q(1,\delta)>0$, for every $\delta \in (0,1]$. 
By the Intermediate Value Theorem, there exists a solution for $Q(W(\delta),\delta)=0$.
Note that
\begin{eqnarray*}
    \frac{\partial Q}{\partial W }  
    & = & 12(\delta + W)^2 - 4W [1+\delta^2+2\delta W] \\
    & \geq & 12(\delta + W)^2 - 4(W+\delta) [1+\delta^2+2\delta W] =  \frac{\partial Q}{\partial \delta},
\end{eqnarray*}
and by substituting $\left[ 1+\delta^2 + 2\delta W \right] = 2(\delta + W)^{3/2}$ we get
\begin{eqnarray*}
    \frac{\partial Q}{\partial \delta} 
    & = & 12(\delta + W)^2 - 4[1+\delta^2+2\delta W](\delta + W) \\
    & = & 12(\delta + W)^2 - 8(\delta + W)^{5/2} \\
    & = & 8(\delta + W)^2(1.5- \sqrt{\delta + W})>0, \ \ \forall (\delta,W)\in(0,1]^2.
\end{eqnarray*}
Hence, both partial derivatives are strictly positive, and the solution $W(\delta)$ to $Q(W,\delta)=0$ is unique.
By the Implicit Function Theorem, we get 
$$
\frac{\partial W(\delta) }{\partial \delta } = - \frac{\frac{\partial Q}{\partial \delta } }{\frac{\partial Q}{\partial W } } <0,
$$
implying that $W= \tfrac{E_2}{E_1 + E_3}$ is decreasing w.r.t.\ $\delta$ in equilibrium.
    \hfill
\end{proof}

\section{Proof of Lemma \ref{Lemma - polarized population}}
\begin{proof}
Note that $\pi_1+ \pi_2+\pi_3 = 2\pi_1+\pi_2 =1$, so it is sufficient to prove that $\frac{\pi_1}{\pi_2}$ is increasing in $\delta$.
Denote $D= \frac{\pi_1}{\pi_2} = \frac{r^*}{\sqrt{\delta +\tfrac{1}{2}r^*}}$ and differentiate with respect to $\delta$, so that
$$
\frac{\partial D}{\partial \delta} = \frac{r^* -\tfrac{dr^*}{d\delta}[2\delta  + \tfrac{1}{2}r^*] }{2(r^*)^2\sqrt{\delta + \tfrac{1}{2}r^*}}.
$$
Since $r^*$ is decreasing in $\delta$, we get $\tfrac{\partial D}{\partial \delta }>0$, and the result holds.
    \hfill
\end{proof}

\end{document}